\documentclass[a4paper,fleqn,usenatbib]{mnras}

\usepackage{newtxtext,newtxmath}
\usepackage{mathptmx}
\usepackage{pdflscape}
\usepackage{amssymb}

\usepackage[T1]{fontenc}
\usepackage{ae,aecompl}

\usepackage{graphicx}	
\usepackage{amsmath}	
\usepackage{amssymb}	
\usepackage{natbib}
\usepackage{hyperref}

\title[]{Chemical modelling of glycolaldehyde and ethylene glycol in star-forming regions} 

\author[A. Coutens et al.]{
A. Coutens,$^{1,2}$\thanks{E-mail: audrey.coutens@u-bordeaux.fr}
S. Viti,$^{1}$
J. M. C. Rawlings,$^{1}$
 M. T. Beltr\'an,$^{3}$ 
 J. Holdship,$^{1}$
 \newauthor
 I. Jim{\'e}nez-Serra,$^{4}$
 D. Qu\'enard,$^{4}$
 and  V. M. Rivilla$^{3}$
\\
$^{1}$ Department of Physics and Astronomy, University College London, Gower St., London, WC1E 6BT, UK \\
$^{2}$ Laboratoire d'Astrophysique de Bordeaux, Univ. Bordeaux, CNRS, B18N, all{\'e}e Geoffroy Saint-Hilaire, 33615 Pessac, France \\
$^{3}$ INAF-Osservatorio Astrofisico di Arcetri, Largo Enrico Fermi 5, 50125 Firenze, Italy \\
$^{4}$ School of Physics and Astronomy, Queen Mary University of London, Mile End Road, London E1 4NS, UK \\
}

\date{Accepted XXX. Received YYY; in original form ZZZ}

\pubyear{2017}

\begin{document}
\label{firstpage}
\pagerange{\pageref{firstpage}--\pageref{lastpage}}
\maketitle

\begin{abstract}
Glycolaldehyde (HOCH$_2$CHO) and ethylene glycol ((CH$_2$OH)$_2$) are two complex organic molecules detected in the hot cores and hot corinos of several star-forming regions.
The ethylene glycol/glycolaldehyde abundance ratio seems to show an increase with the source luminosity. 
In the literature, several surface-chemistry formation mechanisms have been proposed for these two species. With the UCLCHEM chemical code, we explored the different scenarios and compared the predictions for a range of sources of different luminosities with the observations. None of the scenarios reproduce perfectly the trend. A better agreement is, however, found for a formation through recombination of two HCO radicals followed by successive hydrogenations. The reaction between HCO and CH$_2$OH could also contribute to the formation of glycolaldehyde in addition to the hydrogenation pathway. 
The predictions are improved when a trend of decreasing H$_2$ density within the core region with T$\geq$100\,K as a function of luminosity, is included in the model. Destruction reactions of complex organic molecules in the gas phase would also need to be investigated, since they can affect the abundance ratios once the species have desorbed in the warm inner regions of the star-forming regions.
\end{abstract}

\begin{keywords}
astrochemistry -- ISM: molecules -- ISM: abundances -- protostars
\end{keywords}



\section{Introduction}
\label{sect_intro}

Complex organic molecules (COMs) are defined as molecules consisting of 6 atoms or more with at least one carbon atom \citep{herbst2009}.  The warm inner regions of star-forming regions ($>$ 100 K) are known to be enriched in COMs. Their abundant presence in these regions can be explained by the sublimation of the icy grain mantles that release in the gas phase the COMs or the precursors that lead to their formation. These chemically rich regions are called hot cores for high-mass sources and hot corinos for low-mass sources \citep{ceccarelli2006}.
Some of the COMs are particularly interesting because of their potential prebiotic role. This is the case of glycolaldehyde (HOCH$_2$CHO, hereafter GA). This molecule is involved, under terrestrial conditions through the formose reaction, in the formation of ribose, an essential constituent of ribonucleic acid (RNA). It was first detected towards the high-mass star-forming region Sgr B2 by \citet{hollis2000}. The first detection outside of the Galactic Center was obtained towards the hot core G31.41+0.31 by \citet{beltran2009}. It was later detected towards a low-mass protostar by \citet{jorgensen2012} using Atacama Large Millimeter Array (ALMA) science verification data of IRAS 16293-2422. Often studied with glycolaldehyde is its reduced alcohol, ethylene glycol ((CH$_2$OH)$_2$, hereafter EG). The abundance ratio of these two molecules has been determined in several star-forming regions as well as comets to discuss the possible preservation of these species formed at an early stage of the star formation process until their incorporation into asteroids and comets.
The EG/GA ratio is about 4 in comet Lovejoy \citep{biver2015}, $\geq$ 3 in comet Lemmon \citep{biver2014} and $\geq$ 6 in comet Hale-Bopp \citep{crovisier2004}. This ratio is found to be relatively similar in low-mass protostars ($\sim$3--5, \citealt{coutens2015,jorgensen2016}). It is, however, higher for high-mass sources with a value of 10 for G31.41+0.31 \citep{rivilla2017} and lower limits of $\geq$ 6, $\geq$ 13 and $\geq$ 15 in three other sources \citep{lykke2015,brouillet2015}. This led \citet{rivilla2017} to suggest an increase of the EG/GA ratio with the source luminosity (see Table \ref{summary_obs}).

So far, the gas phase routes proposed in the literature do not lead to an efficient formation of glycolaldehyde and ethylene glycol in star-forming regions \citep{woods2012}. Grain surface chemistry is consequently considered as the only way to form these species \citep[e.g.,][]{garrod2008}, which is also supported by laboratory experiments \citep{fedoseev2015,butscher2015,chuang2016}. Their dominant formation pathways are however unclear. A first route (hereafter Scenario 1, see Figure \ref{figure_scenarios}) involving the formation of glyoxal ((HCO)$_2$) through the dimerization of the formyl radical (HCO) followed by its hydrogenation was proposed by \citet{woods2013}, then tested experimentally at low temperature through hydrogenation of CO ices by \citet{fedoseev2015}:
\begin{eqnarray}
\rm HCO + HCO \rightarrow \left(HCO\right)_2, \\
\rm \left(HCO\right)_2  \xrightarrow{2 H} HOCH_2CHO
\end{eqnarray}
\citet{fedoseev2015} also detected ethylene glycol in these experiments and proposed that it formed by hydrogenation of glycolaldehyde:
\begin{eqnarray}
\rm HOCH_2CHO \xrightarrow{2 H} \left(CH_2OH\right)_2
\end{eqnarray}

A second route (hereafter Scenarios 2 and 3, see Figure \ref{figure_scenarios}) based on the recombination of the radicals HCO and CH$_2$OH on the grains was explored in laboratory by \citet{butscher2015} and found to be efficient:
\begin{eqnarray}
\rm CH_2OH + HCO  \rightarrow HOCH_2CHO \\
\rm CH_2OH + CH_2OH  \rightarrow \left(CH_2OH\right)_2
\end{eqnarray}

More recently, \citet{chuang2016} explored the hydrogenation of different CO:H$_2$CO:CH$_3$OH ice mixtures at low temperature and detected both ethylene glycol and glycolaldehyde as well as methyl formate (CH$_3$OCHO, hereafter MF). They found that glycolaldehyde could form by any route (reactions 2 and 4), but that the CH$_2$OH radical recombination reaction (reaction 5) should be less efficient than the hydrogenation of glycolaldehyde (reaction 3).
We will hereafter refer to this case as Scenario 4 (see Figure \ref{figure_scenarios}).

In this paper, we explore the different formation pathways proposed in the literature for these two species by modelling sources of different masses and luminosities.
In Section \ref{sect_model}, we describe the physical and chemical model. In Section \ref{sect_results}, we present the results of the chemical predictions for sources of different luminosities, while in Section \ref{sect_discu} we discuss the results.
Finally, we summarize the results and conclude in Section \ref{sect_conclu}.

\section{The model}
\label{sect_model}

\subsection{The physical and chemical model}

In this study, we perform time-dependent calculations using the gas grain chemistry code UCLCHEM\footnote{\url{https://github.com/uclchem/UCLCHEM}}. This code is fully described in \citet{Holdship2017}.  
To simulate the evolution of the physical conditions during the star formation process, two phases are considered: the free-fall collapse of a cloud (Phase I) followed by a ``warm-up'' phase (Phase II). 

In Phase I, we assume a constant temperature $T$ of 10 K and an increase of the density from an initial value $n_{\rm i}$ = 300 cm$^{-3}$, characteristic of a rather diffuse medium, to a final value $n_{\rm f}$ = 10$^7$ cm$^{-3}$ for high mass sources and $n_{\rm f}$ = 10$^8$ cm$^{-3}$ for low-mass protostars, as assumed in other studies \citep[e.g.][]{woods2013,awad2014}. The initial visual extinction $A_{\rm V}$ is 2 magnitudes and the cosmic ray ionization rate is 1.3\,$\times$\,10$^{-17}$ s$^{-1}$. During this time, atoms and molecules accrete on the grain surfaces with an accretion rate that depends on the density. 
The species may then be quickly hydrogenated or react with other grain species. The initial atomic abundances of He, O, C, and N in our model correspond to the solar values from \citet{asplund2009}, while the other elements (S, Si, Cl, P, F, Mg) are assumed to be depleted by a factor 100.

In Phase II, the density remains constant, while the temperature increases with time from 10 K to 300 K following the trend described in \citet{viti2004}:
\begin{eqnarray}
T\left(t \right) = 10 + a \times t^b.
\end{eqnarray}
This is based on the assumption that the temperature of the gas and dust surrounding the accreting protostar increases according to the same power law as the stellar temperature, and it was fitted so that the maximum temperature of the gas is reached at the contraction time, i.e. the time after which hydrogen starts burning and the star reaches the zero-age main sequence (see more details in \citealt{viti2004}).
The values of $a$ and $b$  derived for hot cores are taken from \citet{viti2004} ($a$ $\sim$ 4.9\,$\times$\,10$^{-2}$, 7.8\,$\times$\,10$^{-3}$, 9.7\,$\times$\,10$^{-4}$, 1.7\,$\times$\,10$^{-4}$, 4.7\,$\times$\,10$^{-7}$ and $b$ $\sim$ 0.63, 0.84, 1.1, 1.3 and 1.98 for masses of 5, 10, 15, 25 and 60 $M_\odot$, respectively) and for a low-mass protostar from \citet{awad2010} ($a$ = 1.9\,$\times$\,10$^{-1}$ and $b$ = 0.53).
This means that the more massive the source, the faster the temperatures raises and reaches its maximum value. The sizes of the hot core ($M$ = 5, 10, 15, 25 and 60 $M_\odot$) and hot corino ($M$ = 1 $M_\odot$) are assumed to be 0.06 pc and 160 au, respectively.
During this phase, the molecules do not freeze any longer and the molecules frozen on the grains can be released in the gas phase by both thermal and non-thermal desorption mechanisms, although above temperatures of 40 K thermal desorption dominates. The thermal evaporation treatment of our model is fully described in \citet{viti2004} and includes monomolecular desorption, volcano desorption and co-desorption with water (see also \citealt{collings2004}). The fractions of methyl formate and glycolaldehyde desorbing due to these different mechanisms were updated based on the temperature programmed desorption experiments carried out by \citet{burke2015}.

\begin{figure}
\begin{center}
\includegraphics[width=0.5\textwidth]{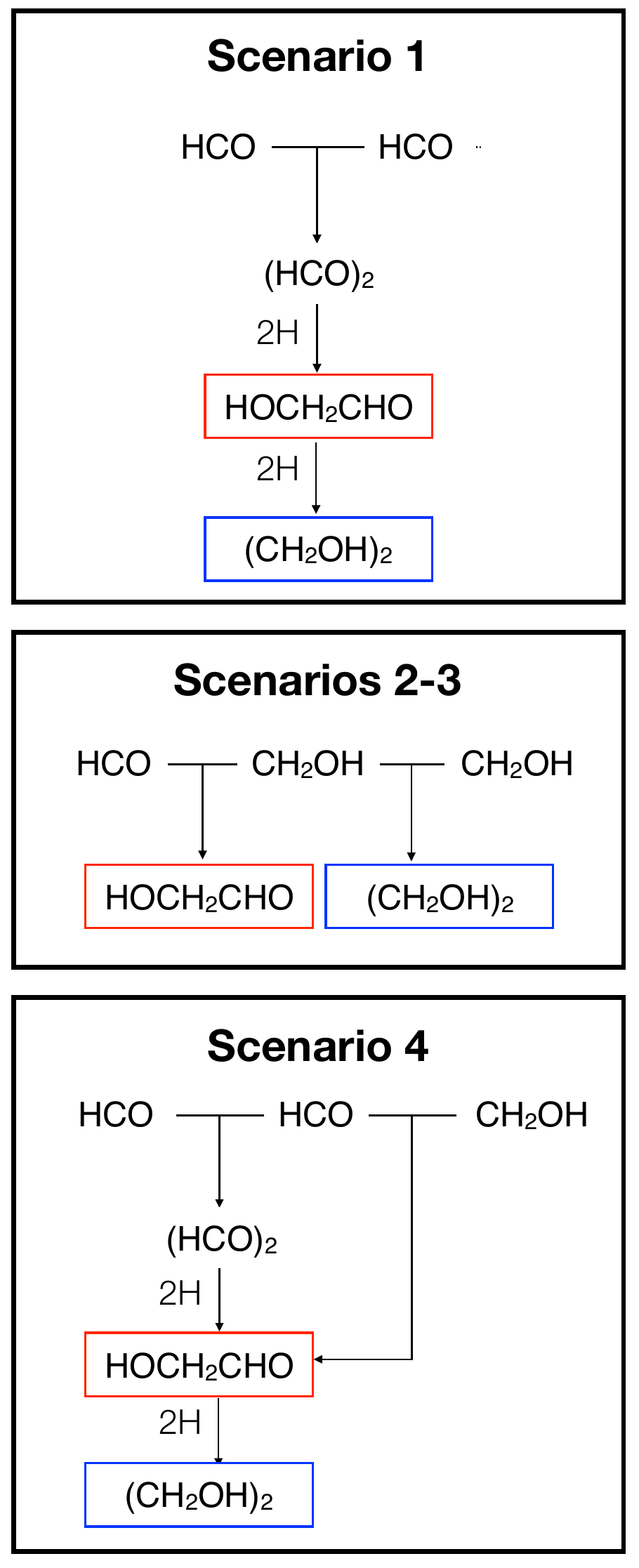}
\caption{Summary of the different scenarios tested in this study. Glycolaldehyde and ethylene glycol are respectively framed in red and blue boxes. } \label{figure_scenarios}
\end{center}
\end{figure}

The gas phase chemistry part of our chemical network is based on the UMIST 2012 database \citep{mcelroy2013}.
Additional gas-phase reactions such as the formation pathway derived by \citet{balucani2015} for methyl formate have also been included.
Our network includes all the potential routes of formation of glycolaldehyde listed in \citet{woods2012,woods2013}, but as stated in these papers, apart from the grain surface reactions (1--2), none of them is efficient. They are consequently not discussed further. For this study, we include all the grain surface routes listed in Section \ref{sect_intro} (see Figure \ref{figure_scenarios}) and vary their rates to test the proposed formation pathways of glycolaldehyde and ethylene glycol.
Some gas-phase destruction routes of glycolaldehyde and ethylene glycol are also included, although we note that the destruction of these two species is very little investigated. In fact the only route of destruction that we found in the literature is the one involving the reaction of glycolaldehyde with OH with a rate of 7.3\,$\times$10$^{-12}$ cm$^{3}$\,s$^{-1}$ \citep{galano2005}. This reaction was consequently added to the network.
We also take into account the dissociation of these molecules by the cosmic ray induced UV field. 
Very few cosmic-ray induced photoreaction rates are included in astrochemical databases due to a lack of data.
For the purpose of this study, we have simply assumed that the rates follow the trends seen in other molecules of
similar types. The rate coefficient of this type of reaction is calculated according to the following formalism: 
$k = \alpha \gamma $/$ \left(1 - \omega \right)$,
where $\alpha$ is the cosmic ray ionisation rate (1.3\,$\times$\,10$^{-17}$ s$^{-1}$), $\gamma$ the scaling factor, and $\omega$ the dust-grain albedo in the far ultraviolet (0.5 at 150 nm).
Thus for glycolaldehyde we adopt a value of
800 for the scaling factor, to be compared with $\sim$1330 and $\sim$520 for formaldehyde
and acetaldehyde, respectively. For ethylene glycol we adopt
a value of 2000, to be compared with $\sim$3050 and $\sim$3500 for methanol
and ethanol, respectively. Our estimates and the dissociation
products are, of course, highly uncertain, but the rates should
be accurate to within a factor of a few.

\subsection{Method}

To test the formation pathways of glycolaldehyde and ethylene glycol, we proceed in two stages. In a first step, we only consider the case of the G31.41+0.3 massive star-forming region, whose mass was found to be about 25 $M_\odot$ \citep{osorio2009}. This source, which is located at a distance of 7.9 kpc, harbors a hot molecular core of about 3\,$\times$10$^5$\,$L_\odot$ rich in complex organic molecules \citep{beltran2005}. We consequently run grids of models by varying the reaction rates for a 25 $M_\odot$ source. We determine the best-fit parameters by comparing the predicted abundances of glycolaldehyde and ethylene glycol with those derived for G31.41+0.31 by \citet{rivilla2017}. 
In a second step, we use these parameters to run models for other source masses (1, 5, 10, 15, and 60 $M_\odot$) and compare the predictions with the observed trend as a function of luminosity (see list of sources in Table \ref{summary_obs}). The relation between the stellar mass and the luminosity used here comes from \citet{molinari2000}. 

To compare the chemical predictions with the observations, we consider that a model is in agreement with the observations, as long as the abundances of the molecules with respect to H$_2$ are reproduced within an uncertainty of one order of magnitude, and that the EG/GA ratio is reproduced within a factor $<$ 2. The large uncertainty for the abundance with respect to H$_2$ is explained by the difficulty to derive precise H$_2$ column densities. If ethylene glycol and glycolaldehyde arise from the same region, their relative abundance ratio is, however, more precise.
A summary of the tested scenarios for the formation paths of glycolaldehyde and ethylene glycol is presented in Figure \ref{figure_scenarios}.
The list of the parameters that are varied are explained below.

First, we consider the fraction of CO that is converted into HCO ($f_{\rm HCO}$) and CH$_2$OH ($f_{\rm CH_2OH}$) on the grain surfaces. It should be noted that this fraction only refers to the fraction of HCO and CH$_2$OH that is left unconverted. More HCO and CH$_2$OH are formed from CO but they are hydrogenated into species such as H$_2$CO and CH$_3$OH. 
We assume that the hydrogenation of CO can lead to the formation of the HCO and CH$_2$OH precursors with a conversion factor of maximum 1\% for the following reasons. In the gas phase of prestellar cores, the abundance of HCO was found to be 10 times lower than CH$_3$OH \citep{bacmann2016}. As the conversion factor of CO into CH$_3$OH is assumed to be $\sim$10\% \citep[e.g.][]{watanabe2002,coutens2017}, we can assume that no more than 1\% of CO should be converted to HCO. It should be noted that in prestellar cores, the formation of HCO could also be due to gas phase formation pathways \citep{bacmann2016}, so this value of 1\% can only be considered as an upper limit. To limit the number of free parameters in this study, we only consider the formation of the precursors HCO and CH$_2$OH on the grains. Note that we can safely do this because the formation routes of glycolaldehyde and ethylene glycol in the gas phase are not efficient, and therefore the only relevant information for our study is the formation of their potential precursors (HCO and CH$_2$OH) on the surface of dust grains. CH$_2$OH has never been observed in the interstellar medium so far; hence we assume a conversion factor similar to the one of HCO, $f_{\rm CH_2OH}$ $\leq$ 1\%.

The other free parameters are the rate coefficients of some key reactions.
Regarding the first scenario (Reactions 1--3), we vary the values of the rate coefficients of the reaction between the two HCO radicals ($k_{\rm HCO+HCO}$) and the hydrogenation reaction of glycolaldehyde ($k_{\rm GA+H}$). The rate coefficient of the HCO + HCO reaction is constant as a function of the temperature. Variations with the temperature cannot affect the results in any case, since, in our model, the hydrogenation reactions including the one for glycolaldehyde only occur during the prestellar phase, i.e. at a temperature of 10 K\footnote{All the results presented in this paper are similar if the hydrogenation reactions occur at temperatures $\leq$ 20\,K.}. It means that glycolaldehyde and ethylene glycol can only form during the first phase, even if glyoxal ((HCO)$_2$) can keep forming at higher temperature in the second phase.
The other hydrogenation reactions are assumed to be very fast, as experimentally found by \citet{fedoseev2015}. 

Regarding the second formation route, we vary the rates of both reactions 4 and 5. We study two cases : i) constant rates ($k_{\rm HCO+CH_2OH}$, $k_{\rm CH_2OH+CH_2OH}$) (scenario 2) and ii) rates increasing with the temperature to simulate the diffusion (by thermal hopping) on the grain surface (scenario 3). In the second case, the rate coefficients are calculated following the formalism described in other studies \citep[e.g.,][]{reboussin2014}. They depend on the density, the number of sites on the surface of the grain per cm$^2$ (1.5 $\times$ 10$^{15}$ cm$^{-2}$), the grain size (0.1 $\mu$m), the binding energy ($E_{\rm b}$) and the mass of the reactants. The diffusion barrier energies ($E_{\rm d}$) are assumed to be equal to 0.5 times the binding energies \citep{garrod2006}. Slightly different factors (0.3--0.7) are also tested to insure that it did not affect the results. The binding energies of HCO (1600 K) and CH$_2$OH (5084 K) are taken from \citet{belloche2014}. 

As the number of free parameters is higher than the number of observational constraints (i.e., abundances of ethylene glycol and glycolaldehyde), we explored several cases with different assumptions : 
\begin{itemize}
\item some cases where reactions can be very fast in order to determine the minimum amount of HCO and CH$_2$OH with respect to CO that is needed to reproduce the observed abundances of glycolaldehyde and ethylene glycol,
\item some other cases where the fraction of CO that is converted to HCO or CH$_2$OH is fixed, 1\% and 0.1\%.
\end{itemize}

\begin{table*}
\begin{center}
\caption{Summary of the scenarios that reproduced the observations of G31.41+0.31 within the uncertainties.}
\label{summary_results}
\begin{tabular}{@{}c@{}cl}
\hline
Scenario & Reactions & Best-fit parameters \\
\hline
1 & $ \rm HCO + HCO \rightarrow \left(HCO \right)_2 $ & A. $ f_{\rm HCO}$ =  0.025\% ; {\bf $k_{\rm \bf HCO+HCO}$ $>$ 10$^{-8}$ cm$^3$\,s$^{-1}$}  ; $k_{\rm GA+H}$ = 2\,$\times$\,10$^{-12}$ cm$^3$\,s$^{-1}$ \\
& $ \rm  \left(HCO \right)_2 \xrightarrow{2 H} HOCH_2CHO$ & B. {\bf $f_{\rm \bf HCO}$ = 1\%} ; $k_{\rm HCO+HCO}$ = 2\,$\times$\,10$^{-15}$ cm$^3$\,s$^{-1}$ ; $k_{\rm GA+H}$ = 8\,$\times$\,10$^{-11}$ cm$^3$\,s$^{-1}$  \\
& $ \rm HOCH_2CHO \xrightarrow{2 H} \left(CH_2OH\right)_2 $ & C. {\bf $f_{\rm \bf HCO}$ = 0.1\%} ; $k_{\rm HCO+HCO}$ =  4\,$\times$\,10$^{-13}$ cm$^3$\,s$^{-1}$ ; $k_{\rm GA+H}$ =  6\,$\times$\,10$^{-11}$ cm$^3$\,s$^{-1}$\\
\hline
2 & $ \rm CH_2OH + HCO  \rightarrow HOCH_2CHO $ & A. $ f_{\rm HCO}$ = 0.001\% ;  $ f_{\rm CH_2OH}$ = 0.03\% ; {\bf $k_{\rm \bf HCO+CH_2OH}$ $>$ 10$^{-8}$ cm$^3$\,s$^{-1}$ ; $k_{\rm \bf CH_2OH+CH_2OH}$ $>$ 10$^{-8}$ cm$^3$\,s$^{-1}$} \\
& $ \rm CH_2OH + CH_2OH  \rightarrow \left(CH_2OH\right)_2 $ & B. {\bf $ f_{\rm \bf HCO}$ = 1\%} ;  $ f_{\rm CH_2OH}$ =  1\% ; $k_{\rm HCO+CH_2OH}$ = 5\,$\times$\,10$^{-17}$ cm$^3$\,s$^{-1}$ ; $k_{\rm CH_2OH+CH_2OH}$ = 7\,$\times$\,10$^{-16}$ cm$^3$\,s$^{-1}$\\
 & (constant rates) & C. {\bf $ f_{\rm \bf HCO}$ = 0.1\%} ;  $ f_{\rm CH_2OH}$ = 0.1\% ; $k_{\rm HCO+CH_2OH}$ = 5\,$\times$\,10$^{-15}$ cm$^3$\,s$^{-1}$ ; $k_{\rm CH_2OH+CH_2OH}$ = 7\,$\times$\,10$^{-14}$ cm$^3$\,s$^{-1}$  \\
\hline
3 & $ \rm CH_2OH + HCO  \rightarrow HOCH_2CHO $ & A. $ f_{\rm HCO}$ = 0.001\% ;  $ f_{\rm CH_2OH}$ =  0.03\% ; {\bf $E_{\rm b}$ (HCO) = 1600 K; $E_{\rm b}$ (CH$_2$OH) = 5084 K ; $E_{\rm d}$ = 0.5 $E_{\rm b}$} \\ 
 & $ \rm CH_2OH + CH_2OH  \rightarrow \left(CH_2OH\right)_2 $ & B. $ f_{\rm HCO}$ = 0.001\% ;  $ f_{\rm CH_2OH}$ =  0.03\% ; {\bf $E_{\rm b}$ (HCO) = 1600 K; $E_{\rm b}$ (CH$_2$OH) = 5084 K ; $E_{\rm d}$ = 0.3 $E_{\rm b}$} \\
& (diffusion) & C. $ f_{\rm HCO}$ = 0.001\% ;  $ f_{\rm CH_2OH}$ =  0.03\% ; {\bf $E_{\rm b}$ (HCO) = 1600 K; $E_{\rm b}$ (CH$_2$OH) = 5084 K ; $E_{\rm d}$ = 0.7 $E_{\rm b}$} \\
\hline
4 & $ \rm HCO + HCO \rightarrow \left(HCO \right)_2$ & A. {\bf $ f_{\rm \bf HCO}$ = 1\% ;  $ f_{\rm \bf CH_2OH}$ = 1\% ; $k_{\rm \bf HCO+HCO}$ =  2\,$\times$\,10$^{-15}$ cm$^3$\,s$^{-1}$ ; $k_{\rm \bf GA+H}$ =  8\,$\times$\,10$^{-11}$ cm$^3$\,s$^{-1}$} ; \\
& $ \rm \left(HCO \right)_2  \xrightarrow{2 H} HOCH_2CHO$ & {\bf ~~~~~$k_{\rm \bf HCO+CH_2OH}$ = 5\,$\times$\,10$^{-17}$ cm$^3$\,s$^{-1}$} \\
& $ \rm HOCH_2CHO \xrightarrow{2 H} \left(CH_2OH\right)_2 $ & B. {\bf $ f_{\rm \bf HCO}$ = 0.1\% ;  $ f_{\rm \bf CH_2OH}$ = 0.1\% ; $k_{\rm \bf HCO+HCO}$ =  4\,$\times$\,10$^{-13}$ cm$^3$\,s$^{-1}$ ; $k_{\rm \bf GA+H}$ =  6\,$\times$\,10$^{-11}$ cm$^3$\,s$^{-1}$ ;} \\
& $ \rm CH_2OH + HCO  \rightarrow HOCH_2CHO$ & {\bf ~~~~~$k_{\rm \bf HCO+CH_2OH}$ = 5\,$\times$\,10$^{-15}$ cm$^3$\,s$^{-1}$} \\
\hline
\end{tabular}
\end{center}
{\bf Note:} The parameters that are fixed in each scenario are indicated in boldface (see more details in the text).
\end{table*}%

\section{Results}
\label{sect_results}

\subsection{Scenario 1}

We explore here the formation of glycolaldehyde and ethylene glycol through the recombination of the HCO radicals followed by successive hydrogenations (Reactions 1--3) in the case of a star of 25 $M_\odot$ and compare the predicted abundances with the ones derived by \citet{rivilla2017} towards G31.41+0.31.

\begin{figure*}
\begin{center}
\includegraphics[width=1.0\textwidth]{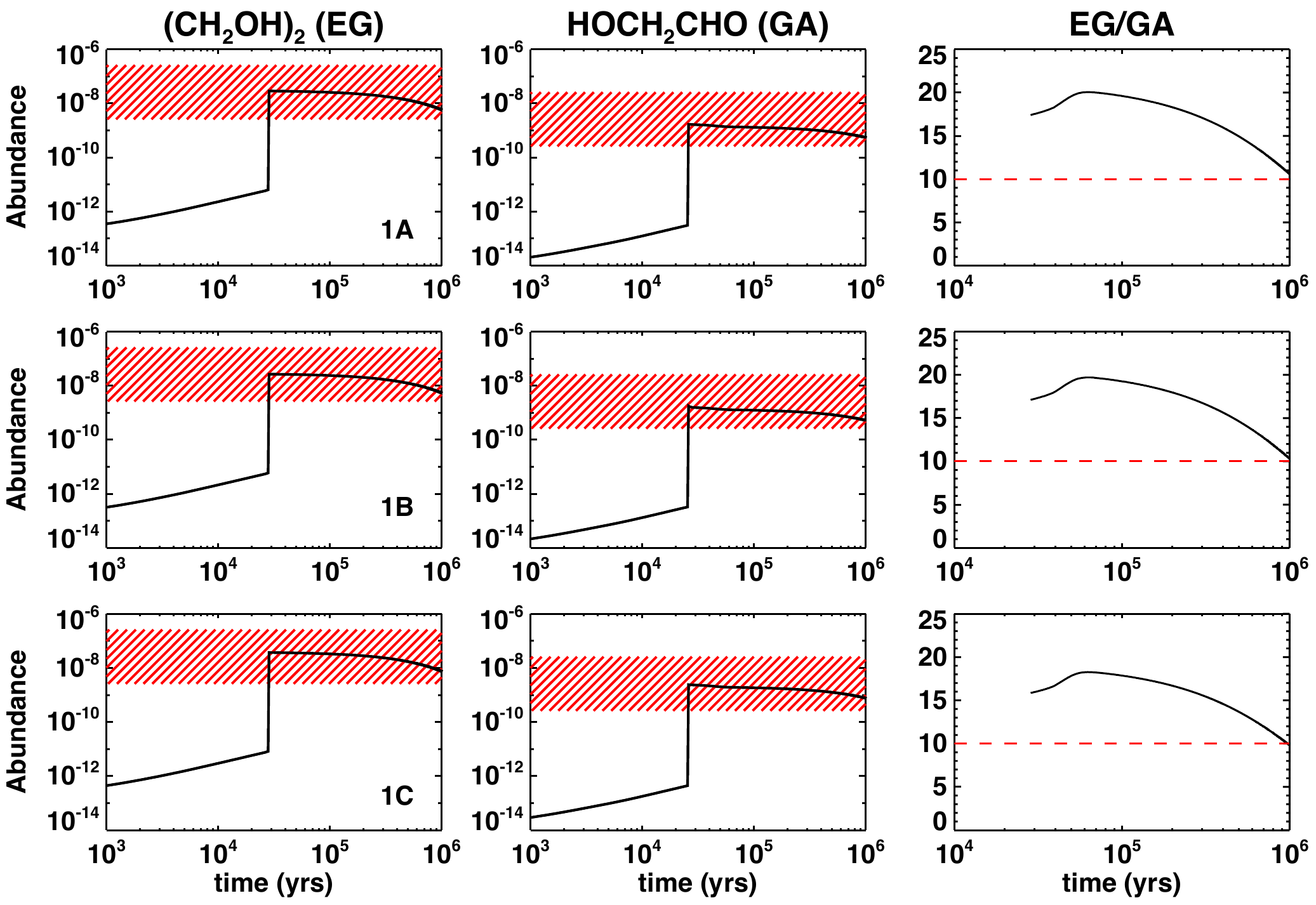}
\caption{Evolution of the abundance of ethylene glycol and glycolaldehyde as a function of time in Phase II in the case of a formation through HCO + HCO reaction followed by hydrogenation (scenario 1). On the left and middle panels, the abundances of ethylene glycol and glycolaldehyde are with respect to H$_2$. The red hatched area indicates the abundances derived for G31.41+0.31 within a factor 10 uncertainty. On the right panel, the EG/GA ratio is only plotted once the two species have totally desorbed. The red dashed line indicates the EG/GA ratio derived for G31.41+0.31. The model number is indicated in the bottom right corner of the left panel (see Table \ref{summary_results}).}
\label{evolution_time_route1}
\end{center}
\end{figure*}

Figure \ref{evolution_time_route1} shows the variation of the abundances of glycolaldehyde and ethylene glycol as a function of time for three different models that give a good agreement with the observations (see Table~\ref{summary_results} for the assumptions of each model).  
 As expected, the gas-phase abundances of glycolaldehyde and ethylene glycol reach a maximum once the two species co-desorb with water at high temperatures ($T$ $\sim$ 100\,K, $t$ = 2--3\,$\times$\,10$^4$ yrs). Afterwards, the abundances decrease slowly. As the age of the source is quite uncertain, we consider that the model is good if the predicted abundance at the time of desorption reaches the observed value. It allows us to determine the minimum fraction of CO that needs to convert into HCO to reproduce the observed abundances. Moreover, at the final time of our calculations $t$ = 10$^6$ yrs, the abundances of the two molecules are still within a factor 10 uncertainty. The EG/GA ratio at the time of the co-desorption with water reflects the EG/GA ratio on the grains. Afterwards, it slightly increases before decreasing. For the models selected here, an EG/GA ratio of $\sim$10 is obtained at $t$ $\sim$ 10$^6$ yrs (see Figure \ref{evolution_time_route1}). This was chosen as a conservative limit. Indeed in most of the high-mass sources (apart from G31.41+0.3), the EG/GA ratio is  constrained by a significantly high lower limit (> 13-15) and, as we realized after testing different source masses that the predicted EG/GA ratio with time was very similar for the high mass sources (see Section 3.4), it is better to have a slightly higher value than 10 at shorter times (3\,$\times$\,10$^4$ < $t$ < 10$^6$ yrs) in order to reproduce the observations of a maximum of sources.
In addition, it should be noted that by using a value of about 10 at $t$ $\sim$ 10$^6$ yrs, the EG/GA ratio of G31.41+0.31 is still consistent with the observations within a factor 2 at any time after the desorption of the grain mantles.  
 
For a very efficient HCO + HCO reaction (scenario 1A), we found that the hydrogenation of CO into HCO on grains needs to have an efficiency of at least $ f_{\rm HCO}$ = 0.025\%. This scenario implies that all the HCO on the grains is converted into glycolaldehyde and ethylene glycol ($\sim$ 10\% and 90\% respectively). In other words, 2.3\,$\times$\,10$^{-4}$ of the CO on the grains convert into ethylene glycol and 2\,$\times$\,10$^{-5}$ into glycolaldehyde. 
If the conversion factor of CO into HCO is higher ($f_{\rm HCO}$ = 1\%), the rate of the reaction HCO + HCO only needs to be about 2\,$\times$\,10$^{-15}$ cm$^3$\,s$^{-1}$, while it is a little higher ($\sim$ 4\,$\times$\,10$^{-13}$ cm$^3$\,s$^{-1}$) if $ f_{\rm HCO}$ is equal to 0.1\%. In the first case ($f_{\rm HCO}$ = 1\%), $\sim$1\% and $\sim$0.1\% of solid HCO are converted into ethylene glycol and glycolaldehyde, respectively. In the second case ($f_{\rm HCO}$ = 0.1\%), $\sim$10\% and $\sim$1\% of HCO are converted. In terms of conversion of the solid CO, this is equivalent, in both cases, to a conversion factor of about 10$^{-4}$ and 10$^{-5}$, respectively.

\subsection{Scenarios 2--3}

The formation of glycolaldehyde and ethylene glycol through the radical-radical reactions HCO + CH$_2$OH and CH$_2$OH + CH$_2$OH (Reactions 4--5) was also investigated and compared with the observations of G31.41+0.31. 

\begin{figure*}
\begin{center}
\includegraphics[width=1.0\textwidth]{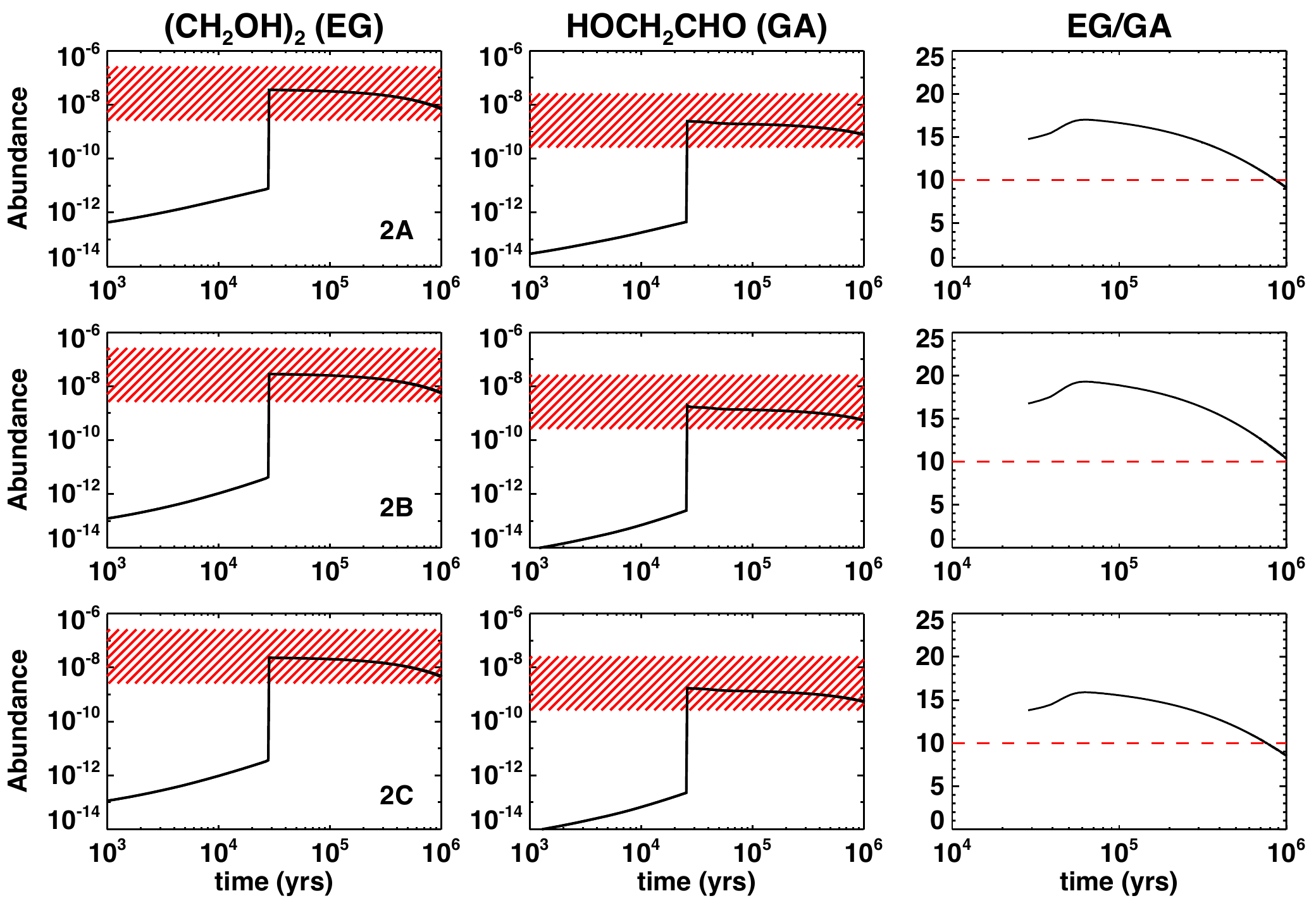}
\caption{Similar to Figure \ref{evolution_time_route1} but for a formation through CH$_2$OH + CH$_2$OH and HCO + CH$_2$OH reactions with constant rates as a function of temperature (scenario 2).}
\label{evolution_time_route2}
\end{center}
\end{figure*}

\begin{figure*}
\begin{center}
\includegraphics[width=1.0\textwidth]{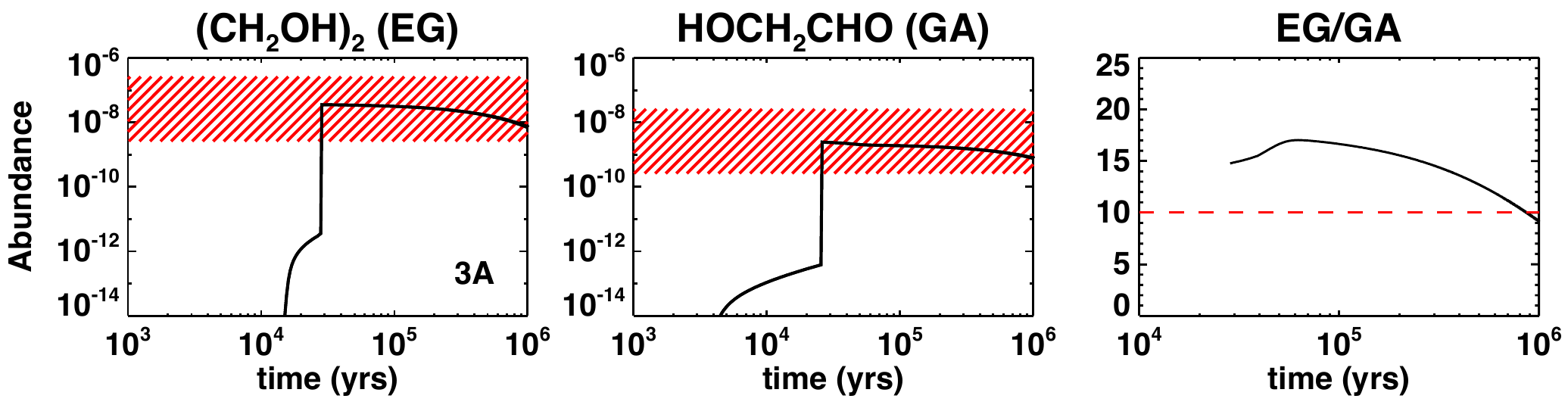}
\caption{Similar to Figure \ref{evolution_time_route1} but for a formation through CH$_2$OH + CH$_2$OH and HCO + CH$_2$OH reactions with diffusion by thermal hopping (scenario 3).}
\label{evolution_time_route3}
\end{center}
\end{figure*}

Figures \ref{evolution_time_route2} and \ref{evolution_time_route3} show the variation of the abundances of glycolaldehyde and ethylene glycol as a function of time in the case of constant rates (Scenario 2) and in the case of diffusion by thermal hopping (Scenario 3) respectively. They are very similar to the ones for route 1.
The minimum fraction of CO that needs to be converted into HCO and CH$_2$OH is estimated to be about 0.001\% (25 times lower than in Scenario 1) and 0.03\% respectively, both in the case of constant rates as a function of the temperature and in the case of diffusion. The only difference is that, in the first case, the formation of ethylene glycol and glycolaldehyde on grains already occur during Phase 1, while in the second case they only form in Phase 2. This explains the lower abundances of ethylene glycol and glycolaldehyde predicted with the diffusion scenario at early times in Phase 2 ($\sim$ 10$^3$--10$^4$ yrs). The non-thermal mechanisms can only release them once they are formed on the grains. It should be noted that for diffusion, assuming a diffusion barrier energy equal to 0.3 or 0.7 times the binding energy (instead of 0.5) does not affect the results (these two cases are consequently not shown in Figure \ref{evolution_time_route3}). The diffusion on the grains starts being efficient at a higher or lower temperature depending on the assumed diffusion barrier energy, but the final abundances are similar.
In the end, 100\% of HCO and 3\% of CH$_2$OH convert into glycolaldehyde, the remaining 97\% of CH$_2$OH are used to form ethylene glycol. It means that about 2\,$\times$\,10$^{-5}$ of CO is converted into glycolaldehyde and 3\,$\times$\,10$^{-4}$ into ethylene glycol.

 In the case of conversion factors of CO into HCO and CH$_2$OH of 1\%, the constant rate of the reaction HCO + CH$_2$OH needs to be about 5\,$\times$10$^{-17}$ cm$^3$\,s$^{-1}$, while the rate of the reaction CH$_2$OH + CH$_2$OH should be about 7\,$\times$\,10$^{-16}$ cm$^3$\,s$^{-1}$. This represents a conversion of 0.1\% of HCO and 0.1\% of CH$_2$OH into glycolaldehyde and 2\% into ethylene glycol.
 The reaction rates are higher by two orders of magnitude (5\,$\times$\,10$^{-15}$ cm$^3$\,s$^{-1}$ and 7\,$\times$\,10$^{-14}$ cm$^3$\,s$^{-1}$), when the conversion factors of CO into HCO and CH$_2$OH are 0.1\%. In this case, 1\% of HCO and 1\% of CH$_2$OH are used to form glycolaldehyde, while 20\% of CH$_2$OH is used for ethylene glycol. This means that, here again, about 2\,$\times$\,10$^{-4}$ and 2\,$\times$\,10$^{-5}$ of the CO on the grains lead to the formation of the observed ethylene glycol and glycolaldehyde, respectively. 

\subsection{Scenario 4}

\begin{figure*}
\begin{center}
\includegraphics[width=1.0\textwidth]{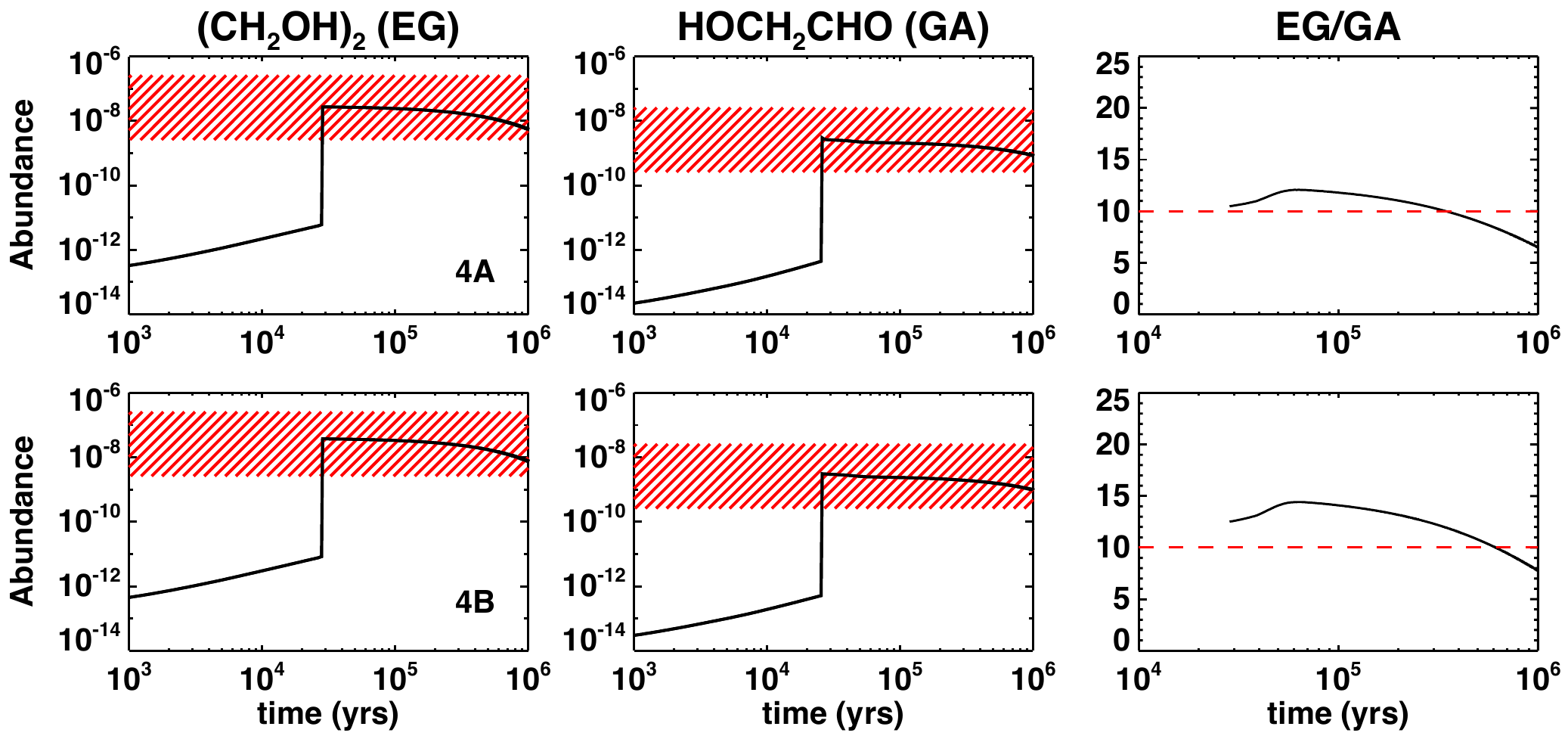}
\caption{Similar to Figure \ref{evolution_time_route1} but for a formation through HCO + HCO reaction followed by hydrogenation and HCO + CH$_2$OH reaction (scenario 4).}
\label{evolution_time_route4}
\end{center}
\end{figure*}

\citet{chuang2016} showed in their experiments that all the routes studied earlier (i.e. a combination of Scenarios 1 and 2) may work, but that the reaction CH$_2$OH + CH$_2$OH (Eq. 5 in this paper) would be less efficient than the other reactions. We tested this scenario and considered that the reaction between the two CH$_2$OH radicals is not efficient at all. This was, moreover, suggested in a theoretical study by \citet{Enrique2016}.
Because of the increase in the number of free parameters, we only ran models with reaction rates ($k_{\rm HCO+HCO}$, $k_{\rm GA+H}$ and $k_{\rm HCO+CH_2OH}$) similar to those determined in the cases 1B/2B ($f_{\rm HCO}$ = $f_{\rm CH_2OH}$ = 1\%) and 1C/2C ($f_{\rm HCO}$ = $f_{\rm CH_2OH}$ = 0.1\%). We found that they can reproduce the observations within the uncertainties, even if the predictions for the EG/GA ratios are slightly lower than those found for the scenarios 1 and 2 separately (see Figure \ref{evolution_time_route4} for comparison with Figures \ref{evolution_time_route1} and \ref{evolution_time_route2}). With these parameters, glycolaldehyde is efficiently formed both through the HCO + HCO pathway and the HCO + CH$_2$OH reaction. In the case of $f_{\rm HCO}$ = $f_{\rm CH_2OH}$ = 1\%, $\sim$0.1\% of HCO and $\sim$0.1\% of CH$_2$OH are converted into glycolaldehyde, while $\sim$1\% of HCO produces ethylene glycol. In the case of $f_{\rm HCO}$ = $f_{\rm CH_2OH}$ = 0.1\%, $\sim$1\% of HCO and $\sim$1\% of CH$_2$OH lead to the formation of glycolaldehyde, and $\sim$10\% of HCO forms ethylene glycol. 

\subsection{Variation of the EG/GA ratio with the luminosity}
\label{sect_mass_variation}

\begin{figure}
\begin{center}
\includegraphics[width=0.45\textwidth]{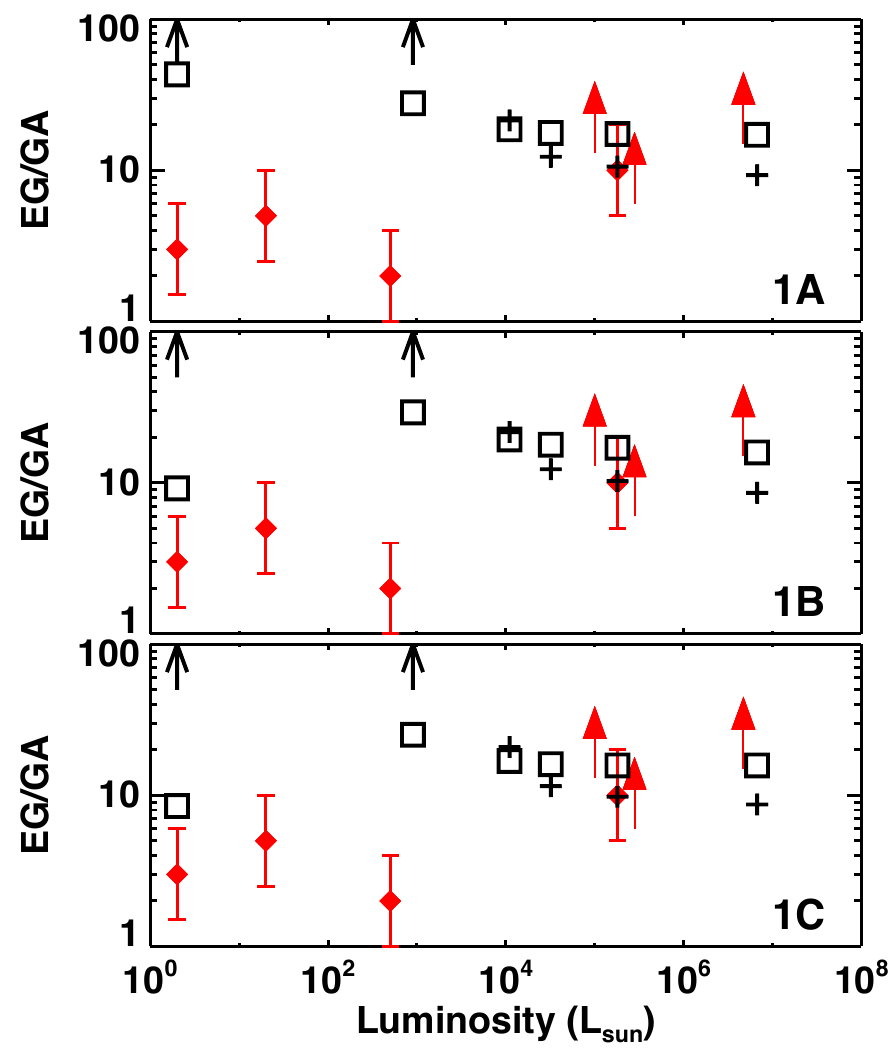}
\caption{Comparison of the EG/GA ratios as a function of the source luminosity in the case of a formation through HCO + HCO reaction followed by hydrogenation (scenario 1). The observations are shown in red: the solid arrows correspond to the lower limits derived in some of the sources, while the diamonds with error bars show the other measurements. The predictions of the EG/GA ratios just after the desorption of the grain mantles and at a time of 10$^6$ years are indicated with black squares and signs ``+", respectively. The cases where the predicted ratio at a time of 10$^6$ years is above 100 are indicated with open black arrows. The model number is indicated in the bottom right corner of the left panel.}
\label{comp_EG_GA_route1}
\end{center}
\end{figure}

\begin{figure}
\begin{center}
\includegraphics[width=0.45\textwidth]{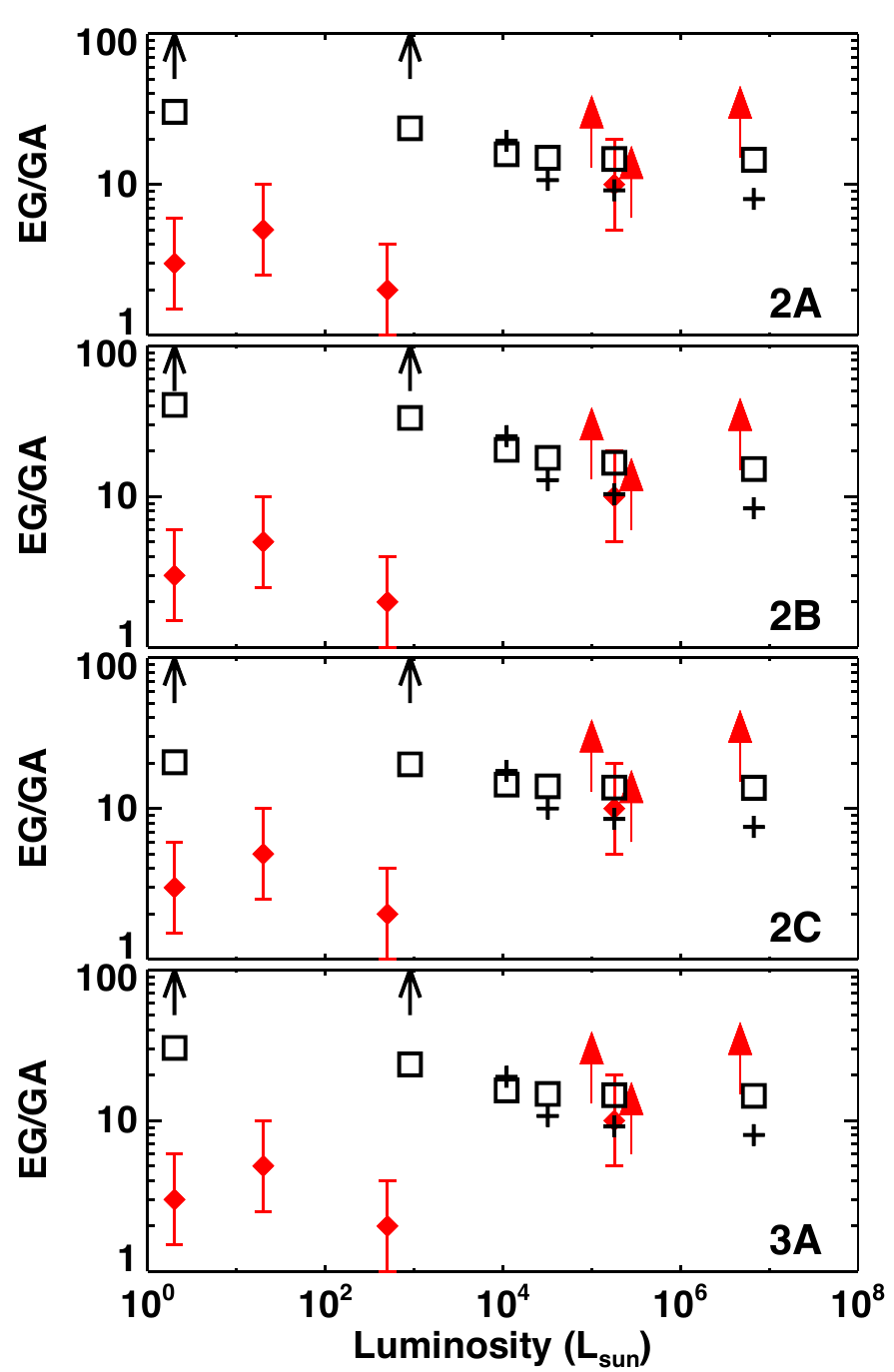}
\caption{Similar to Figure \ref{comp_EG_GA_route1} but for a formation through HCO + CH$_2$OH and CH$_2$OH + CH$_2$OH reactions (scenarios 2-3).}
\label{comp_EG_GA_route2}
\end{center}
\end{figure}

To study the variation of the EG/GA ratio as a function of the luminosity, we run each of the models listed in Table \ref{summary_results} for sources with masses from 1 to 60 $M_\odot$. As explained in Section \ref{sect_model}, the only difference for the intermediate- and high-mass sources is the speed at which the temperature increases depending on the mass, while for the low-mass case, we also increase the density and use a smaller size of 160 au (instead of 0.06 pc for hot cores), which is more characteristic of hot corinos \citep{awad2010}. 
The EG/GA ratios are extracted just after the desorption of the grain mantles and at a time of 10$^6$ years. The values given after the desorption of the grain mantles correspond to the EG/GA ratio inherited from the grain mantles, while the values at a time of 10$^6$ years reflect how the EG/GA ratios are affected by the gas phase destruction routes of ethylene glycol and glycolaldehyde. 
The comparisons of the predicted and observed EG/GA ratios are shown in Figures \ref{comp_EG_GA_route1} and \ref{comp_EG_GA_route2}. 
The EG/GA ratio inherited from the grain mantles appears to be relatively constant (13--20) for the more massive sources with a slight increase for the 5 $M_{\odot}$ objects. The values for the low-mass protostars depend however on the scenario. For models 1B and 1C, the EG/GA ratio is in better agreement with the observations than in the other cases.
At a time of 10$^6$ years, the predicted EG/GA ratio decreases with the luminosity and shows very high values for 1 and 5 $M_\odot$ whatever the scenario is. In fact, the abundance of glycolaldehyde decreases very quickly due to its destruction by OH. The abundance of OH predicted by our models quickly reaches a value of 10$^{-10}$ for the more massive sources, while it can be up to two orders of magnitude higher for sources of 1 and 5 $M_\odot$. 
For this reason, in a low-mass source, glycolaldehyde falls down to an abundance of 10$^{-16}$ in a few 10$^5$ years, which makes this molecule undetectable in hot corinos. As glycolaldehyde is detected towards five solar-type protostars (\citealt{jorgensen2012}, \citealt{coutens2015}, \citealt{taquet2015}, \citealt{desimone2017}), it indicates that the icy grain mantles only desorbed recently in these sources ($t$ $\lesssim$ 2\,$\times$10$^{5}$ yrs). Although we cannot exclude that the OH abundance could be overestimated by our model, the inferred age of these sources tend to be in good agreement with the upper limit we constrained \citep[e.g.,][]{Schoier2002,Webster2003}. 

\begin{figure}
\begin{center}
\includegraphics[width=0.45\textwidth]{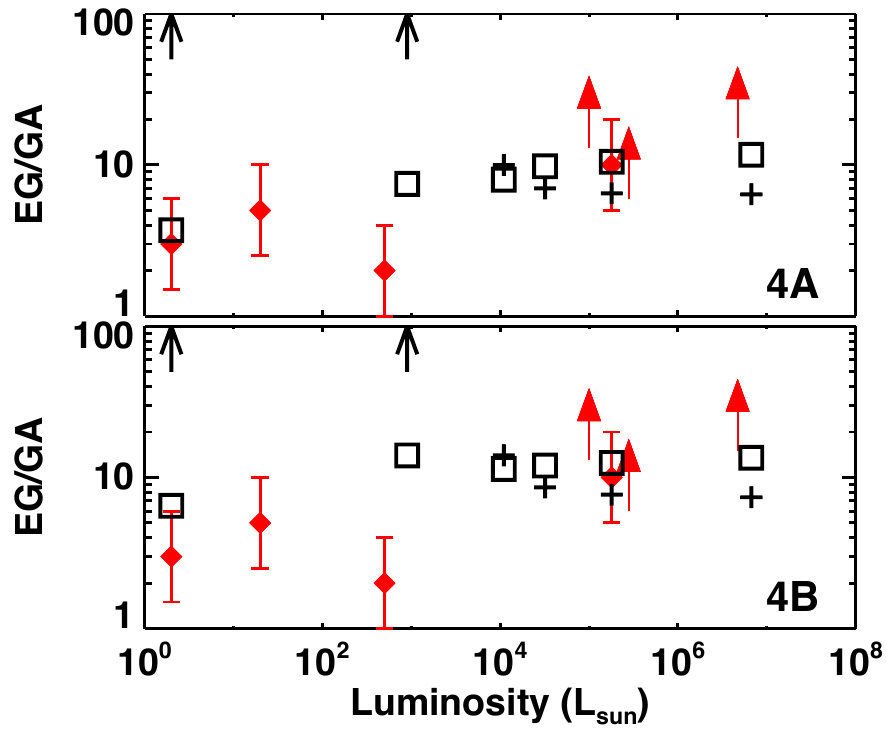}
\caption{Similar to Figure \ref{comp_EG_GA_route1} but for a formation through HCO + HCO reaction followed by hydrogenation and HCO + CH$_2$OH reaction (scenario 4).}
\label{comp_EG_GA_route4}
\end{center}
\end{figure}

The trends for Scenario 4 (for which we did not run any grid) are shown in Figure \ref{comp_EG_GA_route4}. 
The fact that the predicted EG/GA ratios are lower than those obtained within the other scenarios provides a better agreement for low-mass protostars at the time of desorption. They are, however, lower than the lower limits derived in high-mass sources.
But in the end, the trend is relatively similar to what is obtained for models 1B and 1C, in the sense that we see a decrease of the EG/GA ratio for low-mass sources compared to high-mass sources.

\section{Discussion}
\label{sect_discu}

\subsection{Comparison of the observed and predicted trends}

The chemical predictions do not show the same increase of the EG/GA ratios with the luminosity as the observations. This disagreement cannot be explained by the possible different ages of the sources. Indeed the inverse trend is even reinforced with time (see results at $t$ = 10$^6$ yrs). The EG/GA ratio only increases with time for the less massive protostars (leading to predictions above the observed values), while it only decreases for the most massive sources ($M$ = 60 $M_\odot$). 
Two hypotheses can, however, be put forward to explain this disagreement.

First, the gas phase chemical network of glycolaldehyde and ethylene glycol is certainly incomplete. The destruction routes of these two species have been very little explored. 
With the current destruction routes included in our network, we show that the discrepancy between the observed and predicted EG/GA ratios increase with time and the EG/GA ratio is always higher than observed for the low-mass sources. Inclusion of missing routes as well as any difference in the rates of the destruction routes we assumed could potentially lead to the opposite effect. Consequently it appears necessary to constrain the rates of all the possible destruction routes of glycolaldehyde and ethylene glycol for the high temperatures of hot cores and hot corinos ($\sim$ 100--300 K). Astrochemical models of COM formation usually assume that
the environment in which they form is essentially neutral
and chemically saturated. However, if ions and/or radicals
are present then they can act as significant COM destruction
reagents. These reaction channels are not included in most
astrochemical networks and, as an example, we consider the
destruction of GA by OH radicals. Whilst this is possibly
the dominant GA+radical reaction in the conditions that we
are investigating, it should be recognized that there may
well be other important destruction channels that have not
been included in this, and other, studies of COM formation. 

\begin{figure}
\begin{center}
\includegraphics[width=0.45\textwidth]{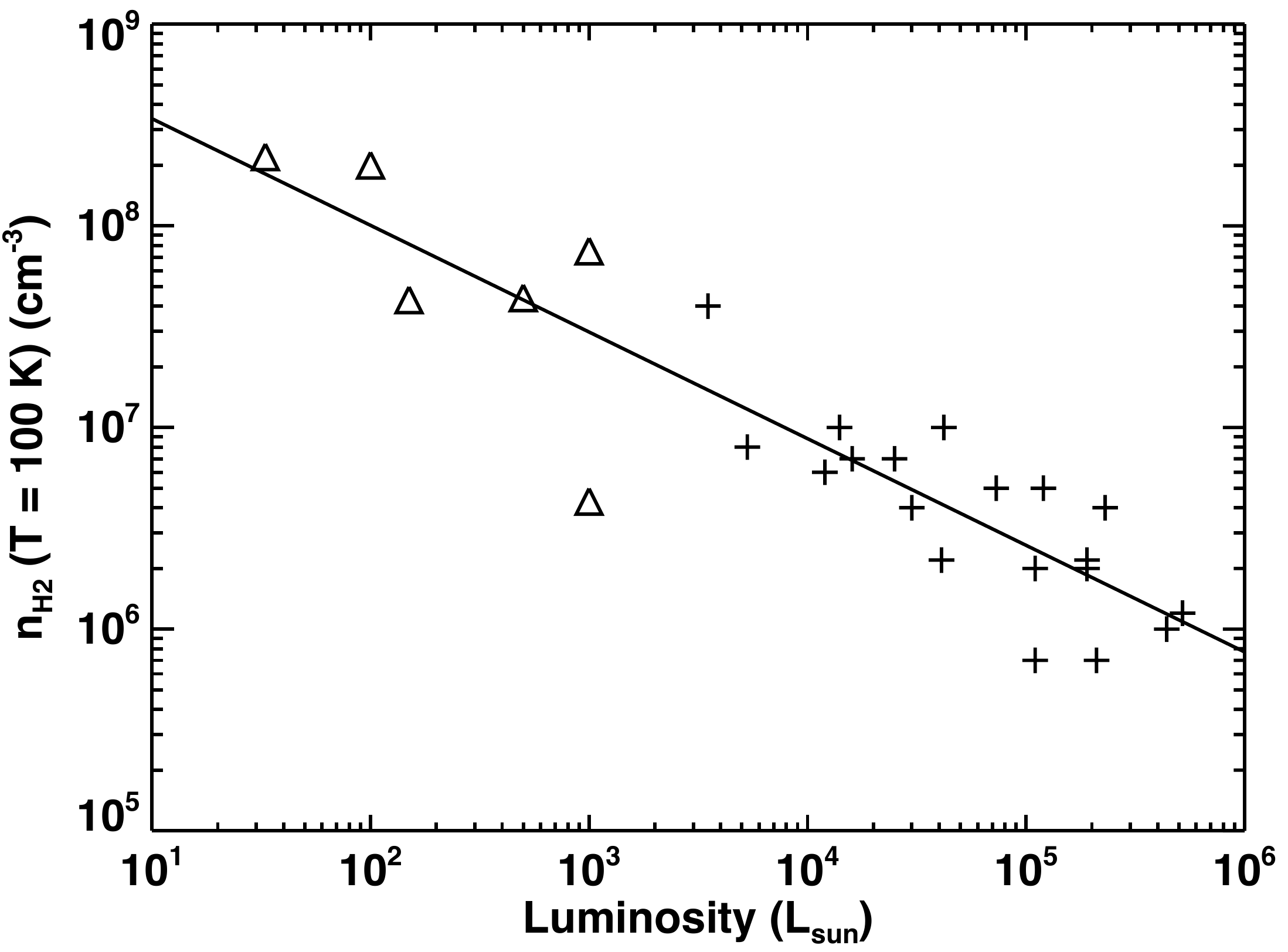}
\includegraphics[width=0.45\textwidth]{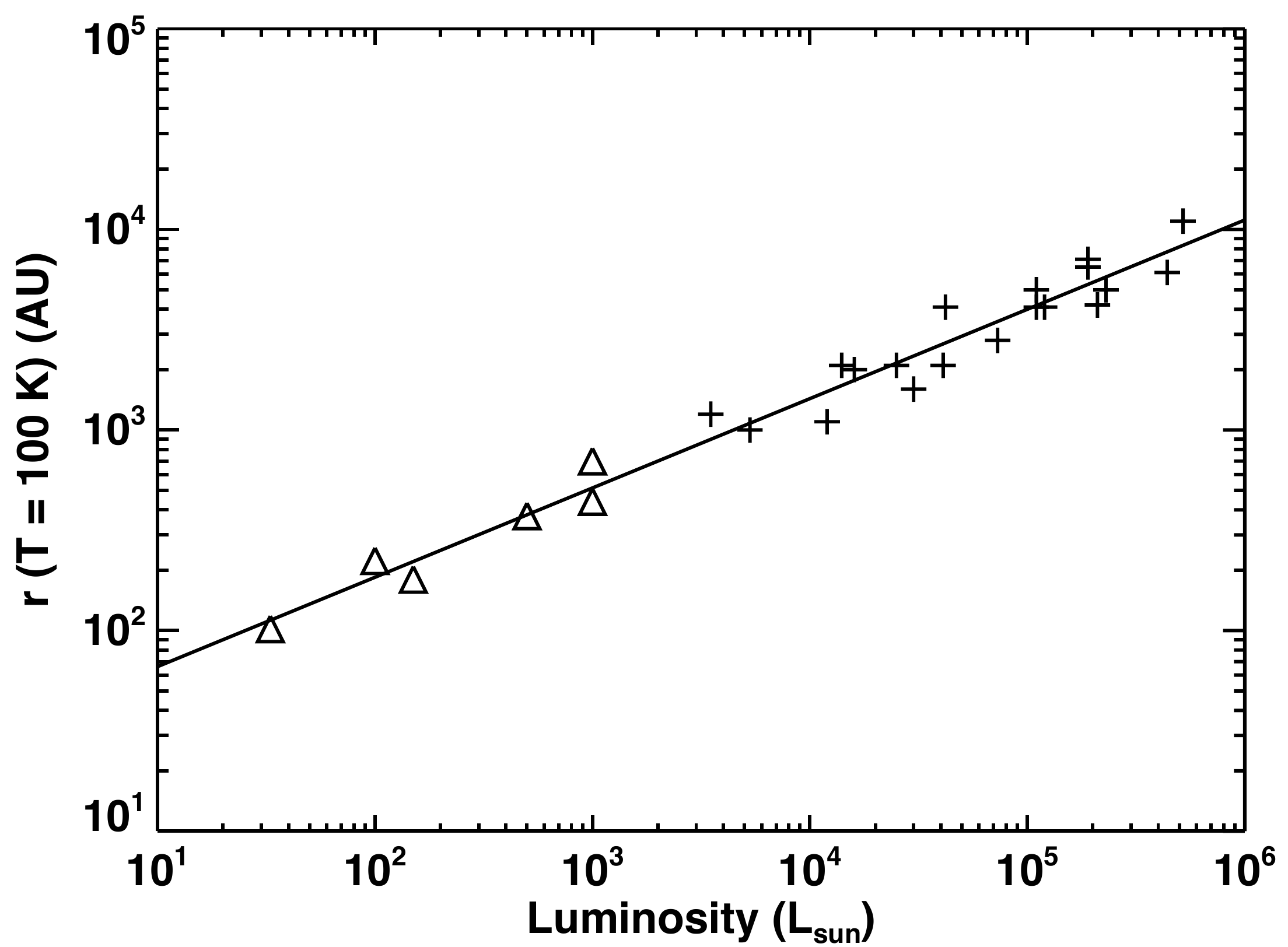}
\caption{Variation of the H$_2$ density (upper panel) and the radius (lower panel) at a temperature of 100\,K in intermediate- and high-mass sources as a function of the luminosity based on the spherical structures determined by \citet{crimier2009,crimier2010} (triangles) and \citet{vandertak2013} (crosses).}
\label{fig_nh2_lum}
\end{center}
\end{figure}

Secondly, some of the physical assumptions may need to be revised. In particular, from Figures \ref{comp_EG_GA_route1} and \ref{comp_EG_GA_route2}, we note that the EG/GA ratios inherited from the grain mantles for low-mass protostars significantly differ according to the scenario, while the ratios for the high-mass sources are relatively similar in all cases. 
A key difference between the chemical models of high mass protostars and those for low mass protostars are the density and size.
To determine if the density and size of hot cores could show any variation with the source luminosity, we explored the literature and plotted the variation of the density of H$_2$ and the radius at $T$=100\,K derived for intermediate- and high-mass sources as a function of their luminosity (see Figure \ref{fig_nh2_lum}). The spherical structures of these sources were constrained by  \citet{crimier2009,crimier2010} and \citet{vandertak2013} based on continuum observations. We find that the density at the position where the temperature reaches 100 K decreases with the source luminosity following the equation:
\begin{equation}
\log \left( n \left({\rm H_2} \right) \diagup {\rm cm}^{-3} \right) = 9.06 - 0.53 \times \log \left(L \diagup L_\odot \right).
\end{equation}
The radius at $T$=100\,K shows a clear increase with the luminosity for both the intermediate and high-mass sources with a best fit : 
\begin{equation}
\log \left( r \left({\rm H_2} \right) \diagup {\rm AU} \right) =  1.38 +  0.45 \times \log \left(L \diagup L_\odot \right)
\end{equation}

\begin{figure}
\begin{center}
\includegraphics[width=0.5\textwidth]{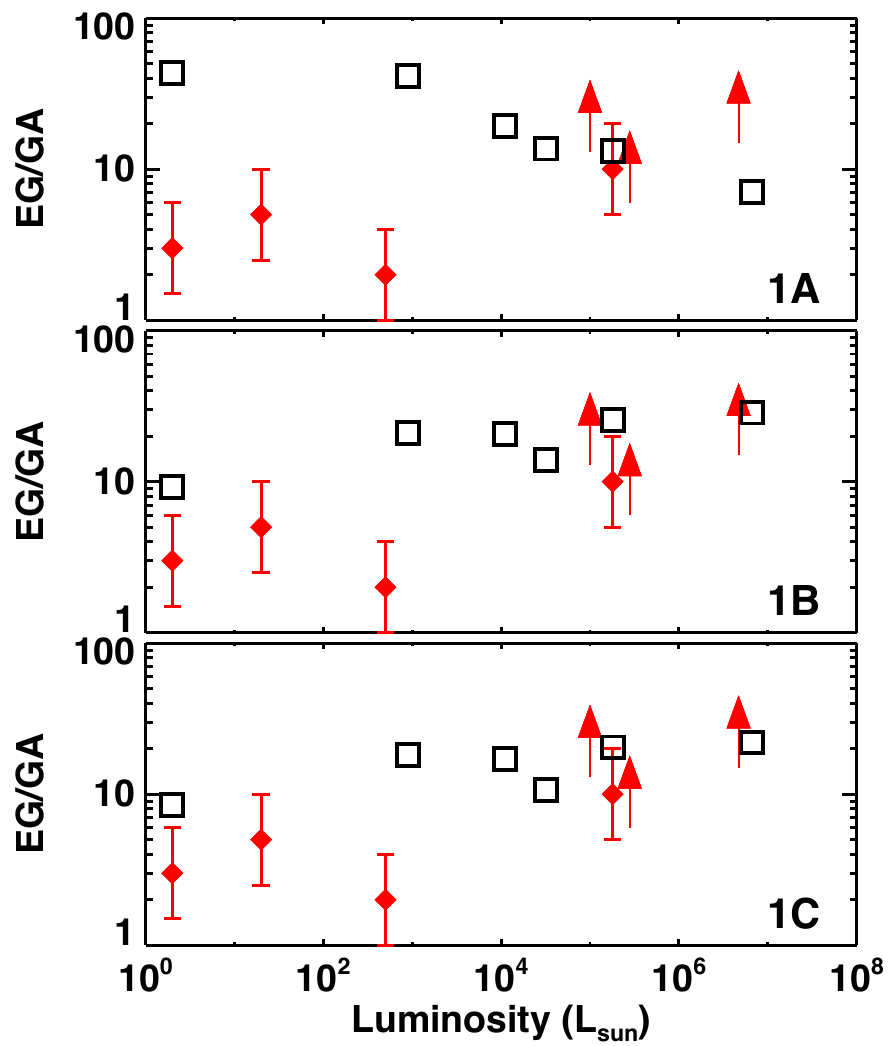}
\caption{Comparison of the EG/GA ratios as a function of the source luminosity in the case of a formation of ethylene glycol and glycolaldehyde through HCO + HCO reaction followed by hydrogenation when a variation of the final density and the size is taken into account (see details in Section \ref{sect_discu}). The observations are shown in red: the solid arrows correspond to the lower limits derived in some of the sources, while the diamonds with error bars show the other measurements. The EG/GA ratios predicted just after the desorption of the grain mantles are indicated with black squares. The model number is indicated in the bottom right corner of each panel.}
\label{comp_EG_GA_route1_dens_size}
\end{center}
\end{figure}

\begin{figure}
\begin{center}
\includegraphics[width=0.5\textwidth]{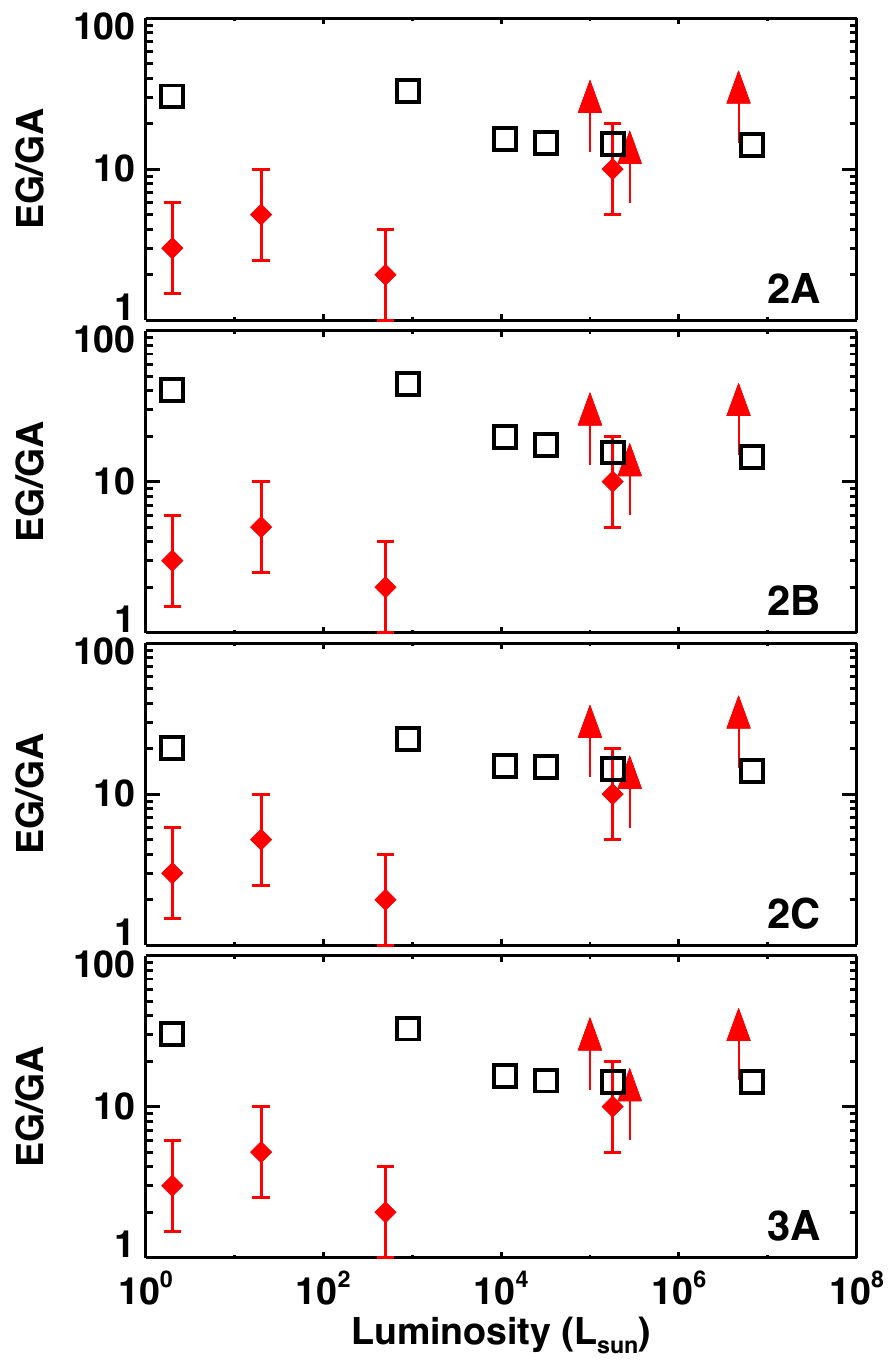}
\caption{Similar to Figure \ref{comp_EG_GA_route1_dens_size} but for a formation of ethylene glycol and glycolaldehyde through HCO + CH$_2$OH and CH$_2$OH + CH$_2$OH reactions (scenarios 2-3).}
\label{comp_EG_GA_route2_dens_size}
\end{center}
\end{figure}

\begin{figure}
\begin{center}
\includegraphics[width=0.5\textwidth]{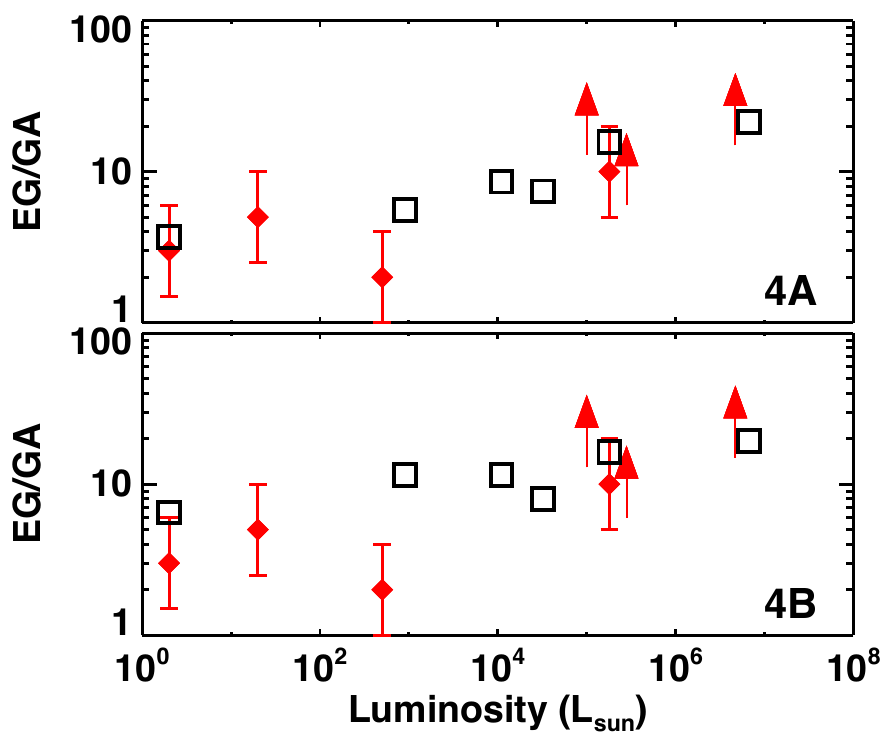}
\caption{Similar to Figure \ref{comp_EG_GA_route1_dens_size} but for a formation through HCO + HCO reaction followed by hydrogenation and HCO + CH$_2$OH reaction (scenario 4).}
\label{comp_EG_GA_route4_dens_size}
\end{center}
\end{figure}

To investigate the possible impact of each of these parameters on the EG/GA ratios, we include the variation of density and radius with the luminosity and re-run the models listed in Table \ref{summary_results} for different luminosities. Figures \ref{comp_EG_GA_route1_dens_size}, \ref{comp_EG_GA_route2_dens_size} and \ref{comp_EG_GA_route4_dens_size} show the new EG/GA ratios predicted after the grain mantle desorption. Changes in the EG/GA ratios are observed. Even if none of the models can perfectly reproduce the trend shown by the observations, the best agreement is found for models 1B, 1C, 4A, and 4B. Note that the density is the parameter that really leads to the change in chemistry. The size has very little impact on the results.
Models 1B, 1C, 4A, and 4B show on average an increase of the EG/GA ratio with the luminosity, which could mean that glycolaldehyde and ethylene glycol will form more efficiently in star-forming regions through HCO + HCO recombination followed by hydrogenation. Glycolaldehyde could, in addition, form through the HCO + CH$_2$OH reaction too. \\

\subsection{Inclusion of methyl formate grain surface formation pathway}
\label{sect_MF}

In all the models presented previously, methyl formate is only formed through gas phase reactions, including the ones proposed by \citet{balucani2015}:
\begin{eqnarray}
\rm CH_3OH + OH  \rightarrow CH_3O + H_2O, \\
\rm CH_3O + CH_3 \rightarrow CH_3OCH_3 + photon, \\
\rm CH_3OCH_3 + F \diagup Cl  \rightarrow CH_3OCH_2 +HF \diagup HCl, \\
\rm CH_3OCH_2 + O  \rightarrow CH_3OCHO + H.
\end{eqnarray}
Compared to \citet{balucani2015}, we used the rate for the reaction OH + CH$_3$OH that was recently determined by \citet{antinolo2016}. We also assumed a depletion of F and Cl by a factor 100 with respect to the solar value. We find that the abundance of methyl formate (with respect to H$_2$) in the high temperature regime (100--300 K) varies between $\sim$7\,$\times$\,10$^{-10}$ and $\sim$2.5\,$\times$\,10$^{-8}$ once the temperature is above 100\,K. \citet{rivilla2017} derived a value of 4.2\,$\times$\,10$^{-8}$ for G31.41+0.31, which is only slightly above the predicted range. Note that even if F and Cl are not depleted, the results are quite similar. The predicted abundance of dimethyl ether, an intermediate species in the formation of methyl formate, is, however, shortly after the thermal desorption of the grain mantles, higher than the observed value ($\sim$ 8.4\,$\times$\,10$^{-8}$, \citealt{rivilla2017}) by one order of magnitude.

Even if our simulations predict abundances of methyl formate relatively similar to the observations with just the inclusion of the gas phase mechanism, it was also shown that methyl formate could also form on the grains with ethylene glycol and glycolaldehyde through hydrogenation of CO:H$_2$CO:CH$_3$OH ice mixtures under certain conditions \citep{chuang2016,chuang2017}. 
Running the previous cases after the inclusion of this grain surface mechanism is beyond the scope of this paper. This would require a higher number of free parameters, since the conversion factor of CO into CH$_3$O and the reaction efficiency between HCO and CH$_3$O are not known either. We should however keep in mind that, if this mechanism is efficient, it implies that some of the free parameters derived in Table \ref{summary_results} could be underestimated. For example, the fraction of CO converted into HCO would need to be higher than the values derived here, since HCO will also react with CH$_3$O to form methyl formate. It may also affect the efficiency of some reactions if they compete between each other.

As an example, we included the grain surface reaction HCO + CH$_3$O and considered the case of diffusion by thermal hopping (scenario 3). The assumed binding energy of CH$_3$O is 2500\,K \citep{garrod2008,belloche2014}. In this case, we only need to vary the fractions of CO converted into HCO, CH$_2$OH, and CH$_3$O.  
We constrain these parameters (see Table \ref{tab_MF}) so that the maximum abundance of methyl formate for a source of 25\,$M_\odot$  reaches a value higher than 4.2\,$\times$\,10$^{-8}$, the precise value of the abundance constrained in G31.41+0.31 by \citet{rivilla2017}. Basically the peak abundance is increased by a factor 2 compared to the case without formation of methyl formate on grain surface, which means that both gas phase and grain surface pathways contribute to the formation of this molecule. We find that $\sim$1\% of CO needs to convert into CH$_3$O to reproduce the observations. The conversion factor of CO into CH$_2$OH is the same as before (0.03\%), while the conversion factor of CO into HCO needs to be higher (0.05\% instead of 0.001\%) as HCO is now used for the formation of both methyl formate and glycolaldehyde. Using these parameters, we also run simulations for a range of 1--60 solar masses and obtain the exact same trend as for the scenario 3A (see Figure \ref{comp_EG_GA_route2}). While the values of the reaction rates and conversion factors of CO into HCO can be affected by the inclusion of the formation pathway of methyl formate on grains, it seems that the predicted trend of the EG/GA ratio with the luminosity is not.

\begin{table*}
\caption{Range of abundances obtained for methyl formate in three different cases}
\label{tab_MF}
\begin{center}
\begin{tabular}{ l c c}
\hline \hline
Assumptions & Range of [CH$_3$OCHO/H$_2$] predicted after desorption \\
\hline
 Gas phase formation of MF only -- F and Cl depleted by a factor 100 & 7\,$\times$\,10$^{-10}$ -- 2.5\,$\times$\,10$^{-8}$ \\
 Gas phase formation of MF only -- no depletion of F and Cl & 7\,$\times$\,10$^{-10}$ -- 2.9\,$\times$\,10$^{-8}$  \\
 Gas phase formation of MF + Scenario 3 (thermal diffusion) & 1.7\,$\times$\,10$^{-8}$ -- 4.7\,$\times$\,10$^{-8}$ \\
  $f_{\rm CH_3O}$ = 1\%, $f_{\rm HCO}$ = 0.05\% and $f_{\rm CH_2OH}$ = 0.03\% & \\
\hline
\end{tabular}
\end{center}
\end{table*}%

\section{Conclusion}
\label{sect_conclu}

In this study, we explored the formation of the complex organic molecules, glycolaldehyde and ethylene glycol. We tested different grain surface formation pathways proposed in the literature by running simulations with the UCLCHEM chemical code. We extracted parameters that give a good agreement between the observations of the source G31.41+0.3 and the chemical predictions for a source of 25\,$M_\odot$. We then ran models for sources from 1 to 60\,$M_\odot$ and compared the variation of the ethylene glycol-glycolaldehyde abundance ratio with the observational trend.
Although none of the scenarios fully reproduce the trend, we found a better agreement with the formation channel involving the recombination of two HCO radicals followed by hydrogenation. It should be noted that a good agreement is also found if glycolaldehyde partially forms through the reaction between HCO and CH$_2$OH in addition to the hydrogenation reaction. 
The reproduction of the trend is improved when a trend of decreasing H$_2$ density within the core region with T$\geq$100 K as a function of luminosity, is included in the model. We also note that destruction reactions of complex organic molecules in the gas phase need to be investigated, as they can affect the abundance ratios after the thermal desorption of the grain mantles in the inner regions of the star-forming regions.

Other species, such as methyl formate, ethanol and dimethyl ether, seem to present abundance ratios that increase or decrease with the luminosity \citep{rivilla2017}. More studies would be needed to understand the origin of these observational trends, as they may provide helpful clues on the formation of these species.

\section*{Acknowledgements}
The work of A.C. was funded by the STFC grant ST/M001334/1 and by the ERC Starting Grant 3DICE (grant agreement 336474). J.H. is funded by an STFC studentship. 
I.J.-S. and D.Q. acknowledge the financial support received from the STFC through an Ernest Rutherford Fellowship and Grant (proposals number ST/L004801 and ST/M004139). V.M.R. acknowledges the funding received from the European Union's Horizon 2020 research and innovation programme under the Marie Sklodowska-Curie grant agreement n. 664931, and from the Italian Ministero dell'Istruzione, Universit\`a e Ricerca through the grant Progetti Premiali 2012 - iALMA.

\bibliographystyle{mnras}
\bibliography{biblio} 



\appendix

\section{Summary of the EG/GA ratios derived in star-forming regions}

\begin{table}
\caption{Summary of the EG/GA ratios derived in star-forming regions}
\begin{center}
\begin{tabular}{@{}c c c c@{}}
\hline \hline
Source & Luminosity & EG/GA & Reference \\
& ($L_\odot$) & \\
\hline
IRAS16293 B & $\leq$ 3 & 3 & \citet{jorgensen2016} \\
NGC1333 IRAS2A & 20 & 5 & \citet{coutens2015} \\
\hline
NGC7129 FIRS2 &  500 & 2 & \citet{fuente2014} \\
\hline
Orion KL & 1\,$\times$\,10$^5$ & $\geq$ 13 & \citet{brouillet2015} \\
G31.41+0.31 & 1.8\,$\times$\,10$^5$ & 10 & \citet{rivilla2017} \\
G34.3+0.2 & 2.8\,$\times$\,10$^5$ & $\geq$ 6 & \citet{lykke2015} \\
W51e2 & 4.7\,$\times$\,10$^6$ & $\geq$ 15 & \citet{lykke2015} \\
\hline
\end{tabular}
\end{center}
\label{summary_obs}
\end{table}%

\bsp	
\label{lastpage}
\end{document}